# Knowledge-Optimising Investment Decisions with Informative Datasets

Sidharth Mallik
*Centre of Excellence for Data Science, Artificial Intelligence and Modelling*
*University of Hull*
Hull, UK
orcid.org/0000-0003-2562-1738

Waymond Rodgers
*Woody L. Hunt College of Business*
*University of Texas at El Paso*
El Paso, USA
orcid.org/0000-0003-4349-5667

Abstract

The enormous growth in datasets, both in number and size, has prompted investors to adapt to new ways for assimilating information. Normatively, the approach has been to integrate such datasets into pricing formulations and assess the performance of portfolios created thereafter. However, such approaches underestimate their influence in portfolio investments by limiting their impact to pricing only. While being theoretically valid, this results in a potential sub-optimal performance in the presence of real-life decision constraints, and a blind spot for performance attribution. We start by analysing investment decisions from a knowledge perspective, which unfurls a new structure. We then propose a FinTech process termed *Knowledge Optimisation* that aims to integrate the influence of knowledge components that could be related to data, models, or business units that extract information. A 3-stage process, namely, decision structure, portfolio selection, and performance assessment is designed. We present an alternative to the *ex-ante* Sharpe Ratio, integrating a term for knowledge units. Through scenario analysis involving portfolio investment situations, we illustrate the utility. By design, the process improves the importance of knowledge in investment decisions.

Keywords: FinTech; Throughput Model; Alternative Data; Knowledge Management; Optimisation

JEL Classification: D81; G11; O32

## 1 Introduction

The modern-day investment ecosystem has witnessed a substantial change arising from large-scale capabilities in data processing and availability. On the one hand, the evolution of the digital economy (1) and the phenomenon of *datification* (5) have made huge amounts of datasets available, while on the other, Data Science modelling techniques have enhanced the capabilities for processing. Integrating such capabilities into investment decisions therefore is an active area for new research. While financial literature categorises such datasets as *Alternative Data*, we identify a subset that we term *Informative Datasets (ID)*, introduced in Definition 1.

Definition 1 (Informative Datasets). *Datasets that contain material information for investments while not representing the price of any instrument.*

The presence of ID introduces complexities in the technological execution of investment decisions, arising from the ability to model such datasets and their availability from 3rd-party sources. An investor faces a situation of *bounded rationality* (23) arising from having limited knowledge of the dataset. Furthermore, the data can be transformed, possibly with an infinite number of mathematical functions. Therefore, justifying their usage configuration to be universally optimal is a difficult problem. In addition, given the 3rd-party sourcing, the same dataset is available to numerous market participants, thereby potentially leading to a form of profit-sharing from the information content, resulting in competition. Therefore, a process that enables optimising investment portfolios in the presence of ID is potentially fruitful in mitigating such concerns and, therefore, reduces the possibility of adverse selection, *ex-ante*.

We consider decision-making as *knowledge work*, and propose a FinTech process termed *Knowledge Optimisation (KO)* that addresses this complexity through a design that optimises for the components involved in investment decisions. We present three important stages of the process. First, the decision is structured with the *Throughput Model (TM)* (18; 20) that integrates ID within the decision mechanism. Second, portfolio optimisation with the Modern Portfolio Theory (MPT) (11) is described as a way to enhance tacit knowledge in the form of a Zero-Knowledge Proof (ZKP) (4). Third, we present a modification of the Sharpe Ratio (SR) (22) that integrates a measure called Return on Knowledge (ROK) (19) that helps assess value-addition. In the next section, we provide a literature review, followed by a methodology that describes KO in greater detail. The mechanism is then illustrated with a scenario analysis. We discuss the robustness, followed by a conclusion on the significance of the process.

## 2 Related Literature

The reviewed literature introduces the key constructs, namely TM, MPT and SR. In addition, the importance of tacit knowledge for investment decisions is presented. To start with a backdrop, investment performance is influenced by knowledge importance of decision inputs. (14) summarises historical perspectives on *Knowledge Management Systems*, citing their importance for attaining competitiveness. The introduction from (6, pp. 1–3) summarises the *heterogeneous expectations hypothesis* whereby in a marketplace, the participants are characterised by varying expectations. The variety in their knowledge reflects through their formation of such expectations. Therefore, knowledge impacts value realised from making investment decisions. To structure this Knowledge-Value relationship, decisions are formed based on TM, described by (16), while (17) summarises the potential for value-addition. Figure 1 introduces TM, where the first part is the core model, and the second part is an improvisation in which an ID acts as the source for information[1]. Examples for ID are available from the context of investing in climate action, such as datasets arising from climate science that are relevant for framing investment decisions. Another source is web datasets such as those available on social media, for which numerous studies exist demonstrating their impact on market prices. A third source is sentiment indicators that have become quite prominent in asset pricing studies recently. To signify the non-price characteristic in Definition 1, alternatives in the form of a price dataset also exist, such as a carbon price, which represents the social cost of carbon. The distinction between the two influences the way they are applied. For example, an ID is generally considered an external variable to pricing, whereas the price dataset could be included as an asset by itself. To summarise therefore, the universe of ID is large and significant, reflecting the importance of our research.

[1] Alternate pathways arising from TM to link the various elements are possible.

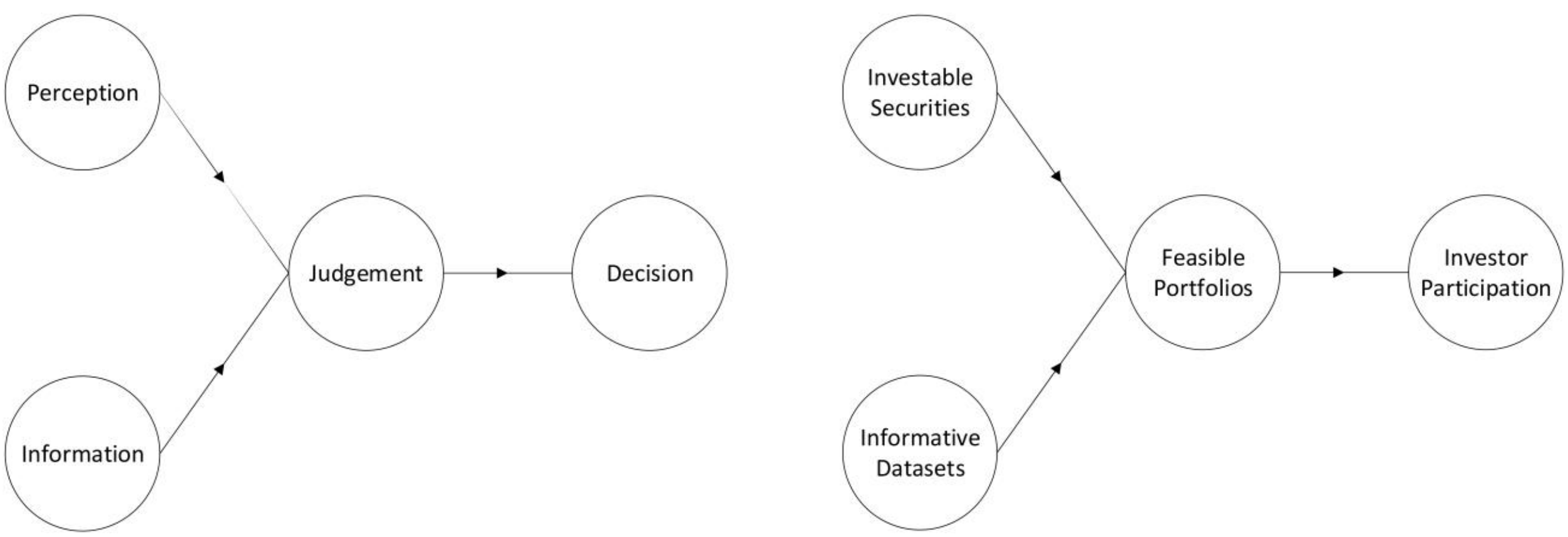

Figure 1: (a) The Throughput Model for Decisions. (b) The Extension with Informative Datasets as the Source for Information.

The integration of ID into portfolio selection can be performed by formulating its importance in the pricing of security returns. In the MPT, the *ex-ante* optimal portfolio is selected by equations (1) and (2),

$$r_p = \sum_{i=1}^{N} w_i \, r_i, \qquad \sum_{i=1}^{N} w_i = 1 \tag{1}$$

$$\min \sigma_{p,t}^2 \quad \text{s.t.} \quad E(r_{p,t}) = r_{target} \tag{2}$$

where $r_p$ is the portfolio return, $w_i$ is the weight assigned to the ith security among N selected ones. Here $r_p$ is represented through moments $\mu_{p,t}$ and $\sigma^2{}_{p,t}$ and $r_{target}$ is the desired target return for the next period. ID is integrated while formulating $r_i$, the security returns, schematically represented in expression (3).

$$r_i \sim F(r_{i,t}, \mu, \sigma, \mathrm{ID}) \tag{3}$$

An example is the GARCH-X formulation (7) where an ID can be integrated in form of an external variable. Performance is assessed with the SR in relation to a benchmark return $r_b$, defined in equations (4) and (5).

$$d = r_p - r_b \tag{4}$$

$$S = \mu_d / \sigma_d \tag{5}$$

where $\mu_d$ and $\sigma_d$ are the *ex-ante* moments of $\tilde{d}$, the estimate for d. Equations (1) to (5) therefore represent the normative approach.

Knowledge components are of two types, codified and tacit (15). Their importance for investment decisions is significant enough for (13, pp. 88–92) to identify them as skills, more specifically as technical skills and tacit skills in investment banking. Our representation of the tacit component as Zero Knowledge allows its material justification through ZKP. To achieve accountability despite a tacit component, evaluating the performance with ROK is considered in equation (6).

$$R_k = r/K \tag{6}$$

where $R_k$ is the ROK for a return r that is realised with $K$ units of knowledge utilised.

# 3 Methodology

The question we face is, when the normative approach introduced in the reviewed literature through equations (1) to (5) already provides a way to assimilate the information, whether there is even a requirement to consider alternatives. We provide a rationale for the gap. The normative way underestimates their effect while making decisions. First, the optimisation equations fail to directly reflect the tacit components of knowledge that the datasets may contribute to. Second, the performance assessment in equation (5) does not explicitly contain the effect from the datasets, instead focusing only on the returns. This makes the approach vulnerable to missing out on certain aspects, such as the underperformance from a 3rd-party dataset. Such a gap could lead to a hidden adverse impact on management performance, becoming a blind spot. Moreover, as the cause is not attributed correctly, measures to improve are hard to identify. Such an effect is more prominent in modern-day markets where the data and the models could both be externally sourced.

To address this gap, we introduce knowledge characterisation for the process that helps unfurl the structure that identifies the importance of the tacit component, as well as leading to modifications of the SR to assimilate the information content of such datasets. The decision structure is represented by the second part of Figure 1 where an ID is integrated in an investment decision through TM. Information in neoclassical finance (21) generally refers to an event or a form of news. We extend the definition to include information represented by datasets in form of ID, commensurate with the perspective from information science (10). Structuring the decision as such is the first stage of KO.

Second is the knowledge perspective on MPT. Markowitz derived the results shown in equations (1) and (2) starting from a set of investable securities. Their moments are estimated from historical returns to arrive at an *ex-ante* optimal set of weights, $w_i$, justifying an economically rational outcome in the face of uncertainty. We consider another perspective for the same approach; one in terms of the way knowledge is applied to selecting portfolios. Initially, the set of investable securities, along with any inputs provided to the optimisation scheme, is observed to be the codified components of knowledge. Before proceeding further, there is no reason to allocate extra weight to any of the securities, thus representing an equi-weighted portfolio[2]. Any mathematical optimisation performed over these components yields a variant that has allocated weights to each security based on its moments. To elaborate, we interpret moments as a numerical representation of knowledge, not a codification of the underlying. Therefore, they can be applied to prove that knowledge exists without unveiling the precise existence. Therefore, such an optimisation enhances performances in presence of knowledge that was not codified. The resulting set of portfolios still contains the same amount of codified knowledge pre-optimisation. They, however, differ in their weight allocation post-optimisation. This difference is observed as tacit knowledge, and the proof of optimality represents a ZKP; in this case the proof represents that the knowledge content is *ex-ante* optimal in face of the observed uncertainty. In other words, the portfolio adds the implicit knowledge that is obtained from moments. The importance of integrating ID is realised when the moments are modelled such as with equation (3) and then fed into the optimisation procedure. They indirectly influence the optimal weights thus adding to the tacit knowledge component. The second stage described here, therefore, unfurls the knowledge structure of portfolio selection.

Having introduced the knowledge perspective, in the third stage, we define a performance measure that aims to assess the importance of the components. The strategy behind the design is that the greater the importance of knowledge, the lower the required number of knowledge units to achieve an equivalent riskadjusted return. We therefore combine equation (5) that defines SR with equation (6) that defines ROK to arrive at equation (7), which we refer to as the modification of SR with knowledge units or $S_k$. The

$$S_k = \mu_d/(\sigma_d * K) \tag{7}$$

risk-adjusted return, or $\mu_d/\sigma_d$, is divided by the number of knowledge units $K$ required to achieve a desired outcome. $K$ could represent a chosen measure of production, depending on how the management decides. For example, where the firm unit that produces the investment strategy is most relevant, then measures such as the size of the unit, or investment amount into the unit are potential candidates. Alternatively, it could represent a trade-off measure that signifies the balance in performance and resources, such as the *cost of compute*. Where the same risk-adjusted return is achieved with a higher $K$, this indicates an underperforming strategy, *ceteris paribus*. This could be occurring due to a depreciation of 3rd-party sourced inputs such as data, model, or both. The measure could also compare the performances of two different investment strategies. The one that achieves the same risk-adjusted return with a lower $K$ is indicative of higher knowledge importance. It includes a new parameter for performance measurement and represents the third stage of KO.

[2] Sometimes referred to as Talmud's Portfolio or equivalent.

# 4 Scenario Analysis

## 4.1 Scenario 1: Rebalancing with Compute

Let's say that a portfolio $P$ is to be rebalanced monthly. At the instant of rebalancing, an interval of size $D$ for revising the positions is identified. Before the next investment, a computational task for recalibrating the decision parameters, given the several constraints in the decision, must be performed. The number of such parameters, in the modern-day context where model complexity is extremely high, could be massive. Even simple Data Science models could have thousands of parameters. High-end industrial-grade models could be several times more. Therefore, the importance of computing during recalibration is hard to ignore, particularly given that the decision is to be made in a bounded time interval $D$. If an investment for carrying out this task is $I$ per interval, and the cost of compute is $C$, then the maximum number of knowledge units $K_{max} = I/C$. The search for optimal parameters must find a solution that is expected to perform at least as well as $r_{target}$, a sufficient threshold of return. If a solution is found before $K_{max}$ units are consumed, then the rebalancing is successful. On the other hand, if none is found, then the investment must be withheld. This places a constraint on real-time performance, even when backtests would have shown that rebalancing was possible in the absence of resource constraints.

## 4.2 Scenario 2: Portfolio Comparison

In this scenario, we present a situation in which applying the two ratios $S$ and $S_k$ leads to conflicting results. Let's say we have two portfolios, $P_1$ and $P_2$ and their corresponding tested results are as mentioned in Table 1 applying the notation from equation (7). $K$ represents a trading desk that can handle a maximum investment amount $I_k$. If $S$ was the guiding measure, then $P_1$ outperforms $P_2$, *ceteris paribus*. However, $P_1$ consumes more knowledge units than $P_2$. This implies that $P_2$ can handle three times the maximum investment. This scalability has implication for gross returns. A $1M invested in $P_1$ gets $100,000 return over the benchmark at an expectation of 10%. However, in $P_2$ the maximum investment can be scaled up to $3M that gets $240,000 at an expectation of 8%. The ratio of gross returns can also be arrived at by dividing the corresponding $S_k$ values for $P_2$ over $P_1$. Thus, the measure adds to performance analysis of investment portfolios and therefore improves portfolio selection.

Table 1: Test Results of Two Portfolios

| | P1 | P2 |
|---|---|---|
| $\mu_d$ | 10% | 8% |
| $\sigma_d$ | 5% | 5% |
| $K$ | 3 | 1 |
| $S$ | 2.0 | 1.6 |
| $S_k$ | 2/3 | 1.6 |

# 5 Robustness and Additional Analysis

The scenarios present the structure that was unfurled with the knowledge perspective. We analyse the importance of the modern-day ecosystem where both data and models could be external. Scenario 1 presents a case that is particularly important for ID. When the data is externally sourced, and many market participants are integrating the same dataset, then greater knowledge importance is required to overcome the competition, and this implies that model complexities might increase, pressurising the knowledge units. Therefore, insufficiency of $K_{max}$ in the case of ID possibly indicates underperformance, negatively impacting valueaddition. Scenario 2 presents an interesting situation for portfolio selection, where the measures are in conflict in the choice of the most suitable. Feeding in the costs related to knowledge units influences the portfolio selection, commensurate with (3).

KO is built upon existing results from MPT and SR as a FinTech process for management involved with investment decisions. The design is generic and therefore could be applied to a wide array of investment strategies, such as buy-and-hold and tracking portfolios. Moreover, the flexibility in the choice allows a range of situations where management decisions could be influenced. The scenarios depict two possible situations, one for deciding on the investment in the face of computational constraints, and the second when scaling with investment turnover is concerned. The former is especially important in the high frequency domain in which decision intervals are small. The FinTech characterisation implies that risks associated with implementing KO are qualified as operational risks, even though the process outcomes are relevant for the decision itself, that falls under decision risk. In this sense, KO has been designed to be positioned for technologically executing investment decisions.

The process is meant to optimise decisions in the presence of both codified and tacit knowledge. The latter presents a dilemma in the sense that an enhancement is claimed in the presence of unobserved knowledge. This is where the concept of ZKP becomes critical in justifying performance. Markowitz showed in a theoretical sense that the portfolios thus arrived are *ex-ante* optimal and this is the same proof we highlight for justifying that an optimality has been reached that includes tacit knowledge. There is a body of literature that argues towards limited applicability of the MPT (8; 12). However, many of the counterarguments arise from either *ex-post* results, or inaccurate moment estimates, both of which while holding significance could be avoided in an *ex-ante* theoretical setting. Our contribution has therefore been to link the proof for MPT to the concept of ZKP that shapes the presented knowledge perspective.

# 6 Conclusion

We have introduced KO as a 3-stage FinTech process that optimises for the knowledge components in investment decisions. We have also presented two scenarios to highlight their importance for realistic situations faced by management. In the modern-day context, such as in approaches with Smart Data (2; 9) which is defined by value and veracity, we have proposed a way for value-addition from knowledge. Moreover, the approach is well-positioned in an ecosystem where data and models could be external and therefore investment performance is vulnerable to competition. The presented scenarios justify their application both in real-time performance and during portfolio selection, signifying utility. Over the course of describing the stages, the two components of codified and tacit knowledge were introduced, and we argue that KO presents an optimal approach in their presence, somewhat counterintuitive when unknowns are involved. Scenario 2 also helps us realise that the scale of value potential with $S_k$ could be multiplicatively higher than with SR, thus highlighting the incentive from utilising KO. The additional term was motivated by the definition of ROK in equation (6), and therefore in conclusion, we assert that KO improves the knowledge importance of investment decisions.